%% file: NIPSIdAdv.tex
\documentclass{article} 
\usepackage{nips12submit_e,times}
\usepackage{amsfonts}      
\usepackage{booktabs} 
\usepackage{array} 
\usepackage{paralist} 
\usepackage{verbatim} 
\usepackage{subfig}
\usepackage{geometry}
\usepackage[nottoc,notlof,notlot]{tocbibind} 
\usepackage[titles,subfigure]{tocloft} 
\usepackage{algorithm}
\usepackage{algorithmic}
\usepackage{color}
\usepackage{graphicx}
\usepackage{verbatim}
\usepackage{hyperref}
\usepackage{amssymb}
\usepackage{amsmath}
\usepackage{cancel}

\newcommand{\beq}{\begin{equation}}
\newcommand{\eeq}{\end{equation}}
\newcommand{\beqa}{\begin{eqnarray}}
\newcommand{\eeqa}{\end{eqnarray}}
\newcommand{\bfig}{\begin{figure}[htb]}

\newcommand{\efig}{\end{figure}}

\newcommand{\mb}[1]{{\mathbf{#1}}}

\newcommand{\ut}{^{(t)}}

\newcommand{\hide}[1]{}
\title{Evaluating Crowdsourcing Participants\\ in the Absence of Ground-Truth}
\author{
Ramanathan Subramanian\\
Northeastern Univ.\\
Boston MA 02215\\
\And
R\'{o}mer Rosales\\
LinkedIn\\
Mountain View CA 94043 
\And
Glenn Fung\\
Siemens\\
Malvern PA 19355\\
\And
Jennifer Dy\\
Northeastern Univ.\\
Boston MA 02215\\
}
\nipsfinalcopy 
\begin{document}
\maketitle
\begin{abstract}
Given a supervised/semi-supervised learning scenario where multiple
annotators are available, we consider the problem of identification of
adversarial or unreliable annotators.
\end{abstract}

\input{intro.tex}

\input{evaluate.tex}

\input{experiments.tex}

\subsection*{References}
\small
\vspace{-0.4em}
[1] V.~C. Raykar et al. Learning from crowds. {\em J. Machine Learning Research}, 11(Apr):1297 - 1322, 2010 
\vspace{-0.3em}

[2] R.~Snow et al. {E}valuating non-expert annotations
  for natural language tasks. In Proc. {\em EM-NLP}, 2008.

\vspace{-0.4em}
[3] Y.~Yan et al. Modeling annotator expertise: Learning when everybody knows a bit of
  something. In Proc. {\em Int. Conference on Artificial Intelligence and Statistics}, 2010.
\vspace{-0.3em}


\end{document}

%% file: intro.tex
\section{Introduction}
Data can be acquired, shared, and processed by an increasingly larger
number of entities, in particular people. The distributed nature of
this phenomenon has contributed to the development of many
crowdsourcing projects. 
This scenario is prevalent in most forms of expert/non-expert
group opinion and rating tasks (including many forms of internet or
on-line user behavior), where a key element is the aggregation of
observations-opinions from multiple sources.

This data acquisition setting has created a number of interesting
problems and opportunities for machine learning and data modeling
among other disciplines. Efficiently aggregating and obtaining
knowledge from multiple distributed sources is far from being a solved
problem. From a machine learning point of view, we treat observations
and opinions as data points and participant label/annotations
respectively. In this sense, multiple participants (annotators) may assign
conflicting labels to the same data point. Clearly,
participant expertise varies and some data points may be considered
{\it easier} for certain participants relative to others. Thus, the
selection of and trust placed on participants plays a crucial role on
crowdsourcing outcomes.

This paper describes an approach for learning to label/annotate by
modeling participant expertise, and focuses on a novel approach to
identify unhelpful or adversarial participants. The paper presents an
approach for participant (also referred to as annotators) evaluation
based on a multi-annotator statistical model and provides experimental
validation of the approach with respect to number of adversaries,
degree of adversarial behavior, and effect on the classification accuracy.



\section{Problem Setup}
Let the available data collected from multiple annotators be given by
a set of points $X=\{ \mathbf{x}_1,\ldots, \mathbf{x}_N \}$ drawn
independently from an input distribution. Some of these data points
may have been labeled by one or more annotators. We denote by $y_i^{(t)}$
the label for the $i$-th data point given by annotator $t$ and let
$Y=\{y_i^{(t)}\}_{it}$. We let $\mathbf{x}_i$ and $y_i\ut$ be random variables in an input and
label space respectively and introduce additional random variables
$Z=\{z_1,\ldots,z_N\}$ to represent the {\it true} but not observed
label for the data points. 

Annotators may label
certain data points more reliably than others because of
the properties of the data point itself. For example, a noisy data
point maybe more difficult to label by all annotators, but also a data
point may be more familiar to certain annotators and thus the
annotator-specific accuracy would vary. Thus, we let the annotation
$y^{(t)}$, provided by labeler $t$, depend on the true unknown label
$z$ but also on the input $\mathbf{x}$.

\begin{figure}[t]
\begin{center}
\includegraphics[width=3.9cm]{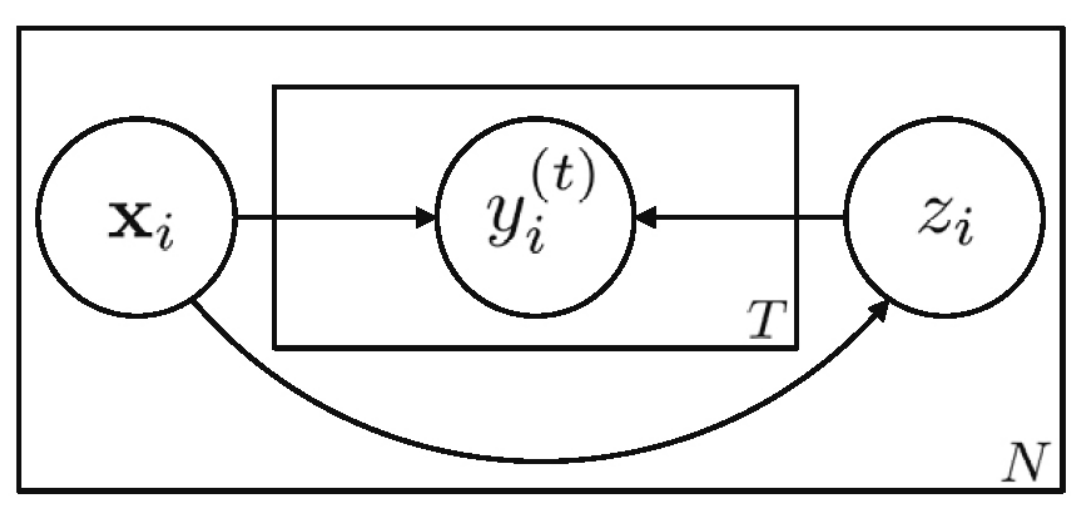}
\hspace{-0.5em}
\includegraphics[width=5.6cm]{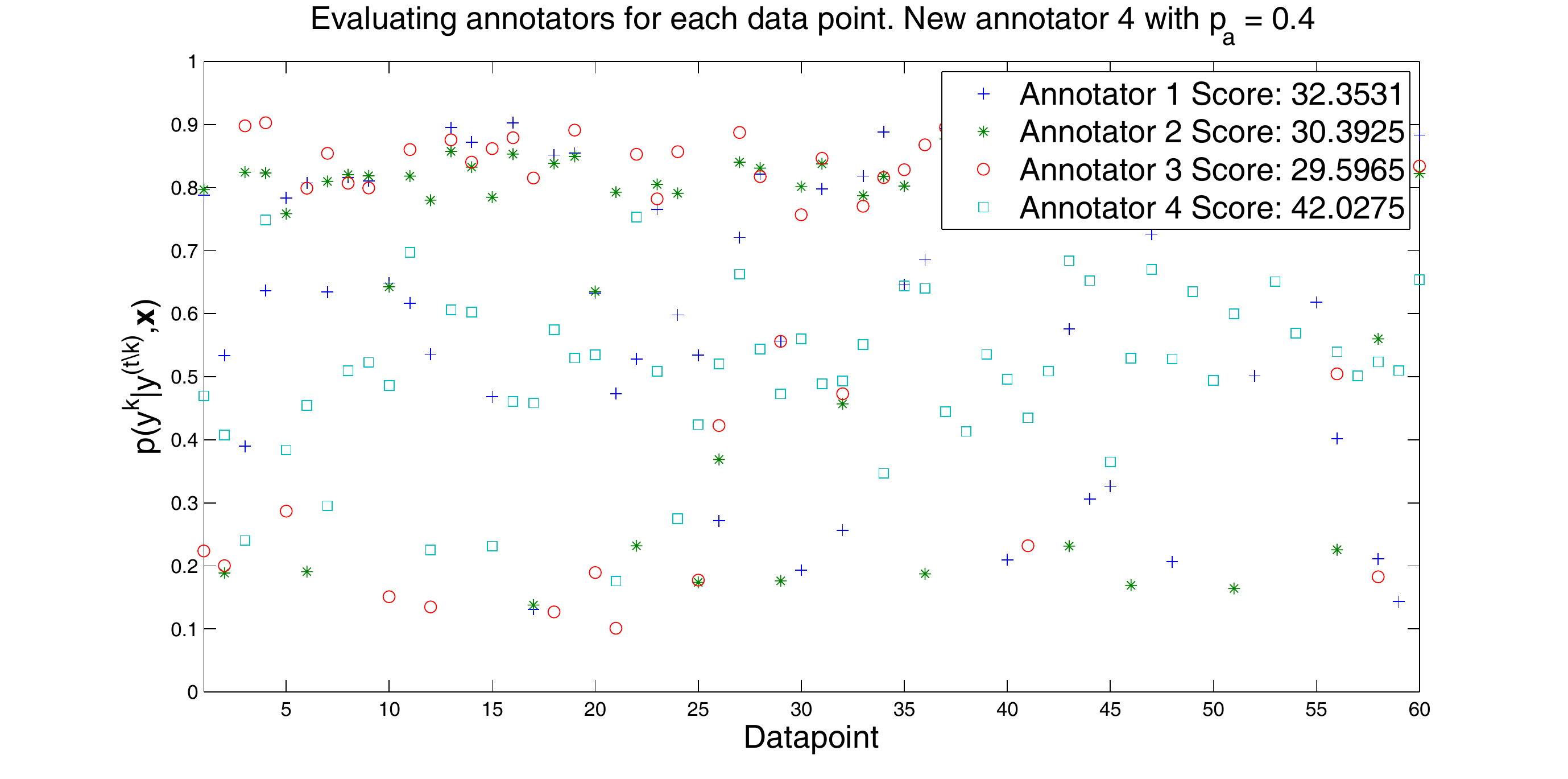}
\hspace{-1em}
\includegraphics[width=5.8cm]{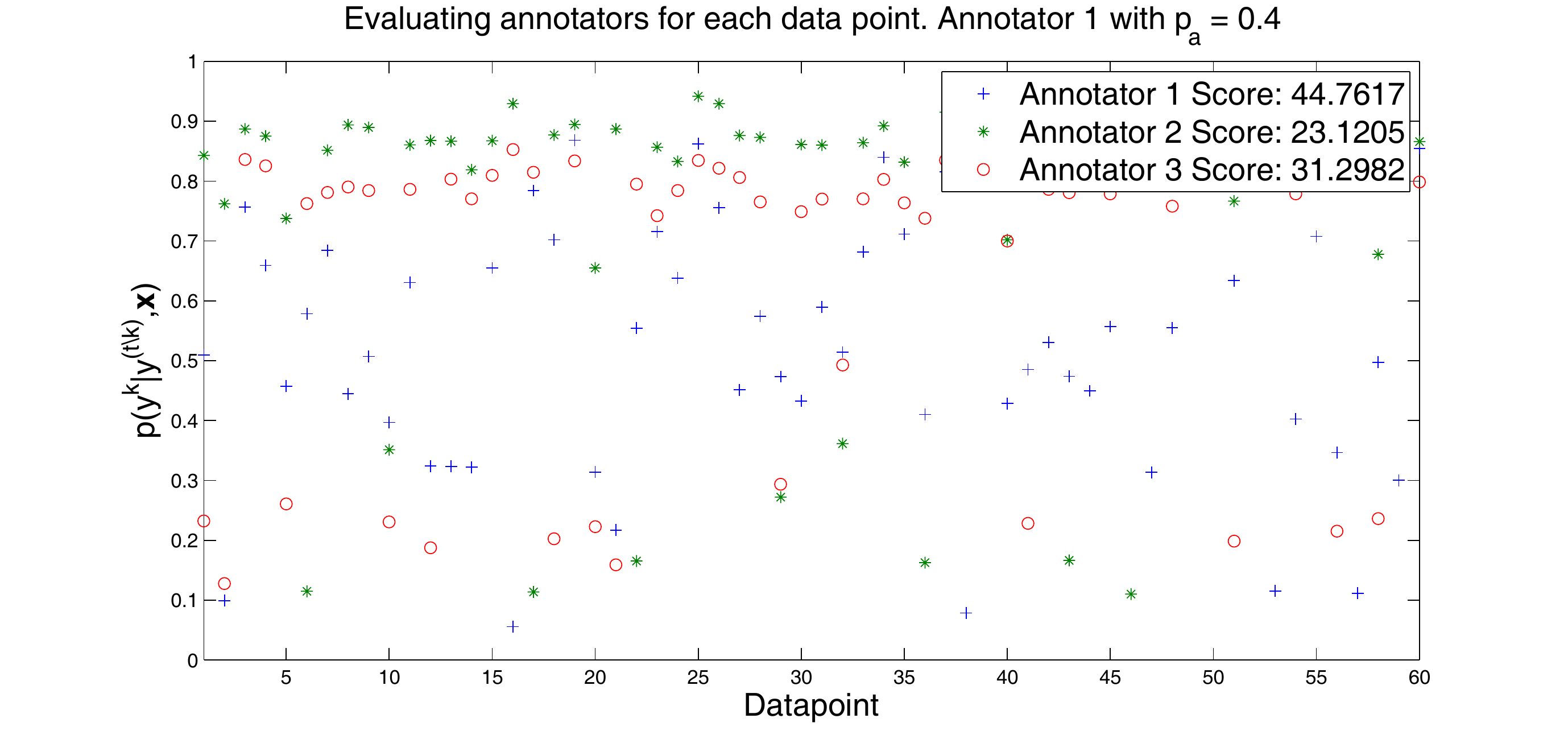}
\caption[9pt]{Left: Graphical Model for $\mathbf{x}$, $\mathbf{y}$, and $z$ respectively inputs, annotator-specific labels, and ground truth label (left). Center-Right: Probability corresponding to Eq.\ref{eq:eval_annot} for each data point: (center) one adversary introduced with  $p_a=0.4$, (right) annotator\_1's labels randomly flipped with probability $p_a=0.4$.}
\label{fig:gm}
\end{center}
\end{figure}

This is in contrast with the related work in [1,2] where annotators are assumed uniformly
accurate/inaccurate irrespective of the data point being labeled.
We instead follow the model introduced in [3] where the
dependency $\mb{x} \rightarrow y^{(t)} \leftarrow z$ is explicit. This
can be summarized by the graphical model in
Fig.~\ref{fig:gm}-Left. 

We focus on binary classification and use a
Bernoulli distribution to model the ability of the annotators to provide the correct label:
$
p(y^{(t)}_i|\mathbf{x}_i,z_i) =  {(1-\eta_t(\mathbf{x}))}^{|y^{(t)}_i-z_i|}{\eta_t(\mathbf{x})}^{1-|y^{(t)}_i-z_i|},
$
where the probability of success $\eta_t(\mathbf{x})$ is a function of $\mb{x}$, thus
allowing the annotator accuracy to depend on the input data point coordinates.
We let the true label depend on $\mb{x}$ via a logistic regression model for $p(z_i=1|\mathbf{x}_i)$.

%% file: evaluate.tex
\section{Evaluating Annotators}
\label{sec:eval_annot}
Consider any single data point. If we knew the ground-truth, then we
could straightforwardly evaluate the annotator accuracy. What if we do
not have access to the ground-truth (it does not exist or is expensive
to obtain)? The following proposed distribution provides a way to
evaluate an annotator without reliance on ground-truth. The basic idea
is to evaluate a conditional distribution measuring how predictable an
annotator label is conditioned on other annotators:
\begin{eqnarray}
\label{eq:eval_annot}
p(y^{(k)}|\{y^{(t \backslash k)}\},\mathbf{x})=\frac{p(\{y^{(t)}\}|\mathbf{x})}{p(\{y^{(t \backslash k)}\}|\mathbf{x})}
=\frac{\sum_z p(\{y^{(t)}\}|z,\mathbf{x})p(z|\mathbf{x})}{\sum_z p(\{y^{(t \backslash k)}\}|z,\mathbf{x})p(z|\mathbf{x})}
\end{eqnarray}

We note that if the ground-truth is given (along with the input data),
the annotators are mutually independent and $p(y^{(k)}|\{y^{(t
  \backslash k)}\},\mathbf{x})=p(y^{(k)}|\mathbf{x})$, as expected. A
related way to understand Eq.~\ref{eq:eval_annot} is as a measure of
how much an annotator's label, for a given data-point, can be
correctly inferred from that of other annotators.

%% file: experiments.tex
\section{Experiments}

\begin{figure}[t]
\begin{center}
\includegraphics[width=4.8cm]{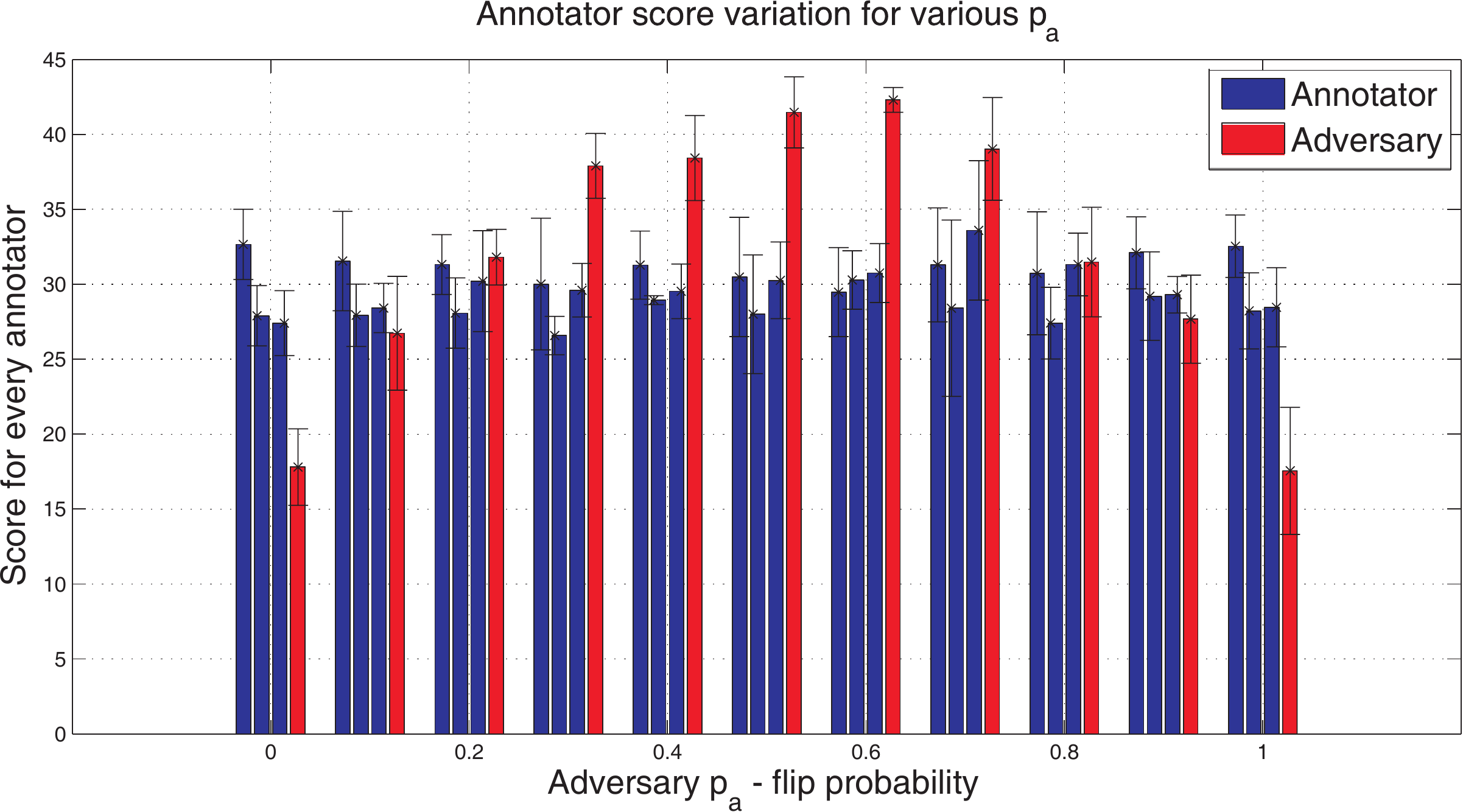}
\includegraphics[width=4.8cm]{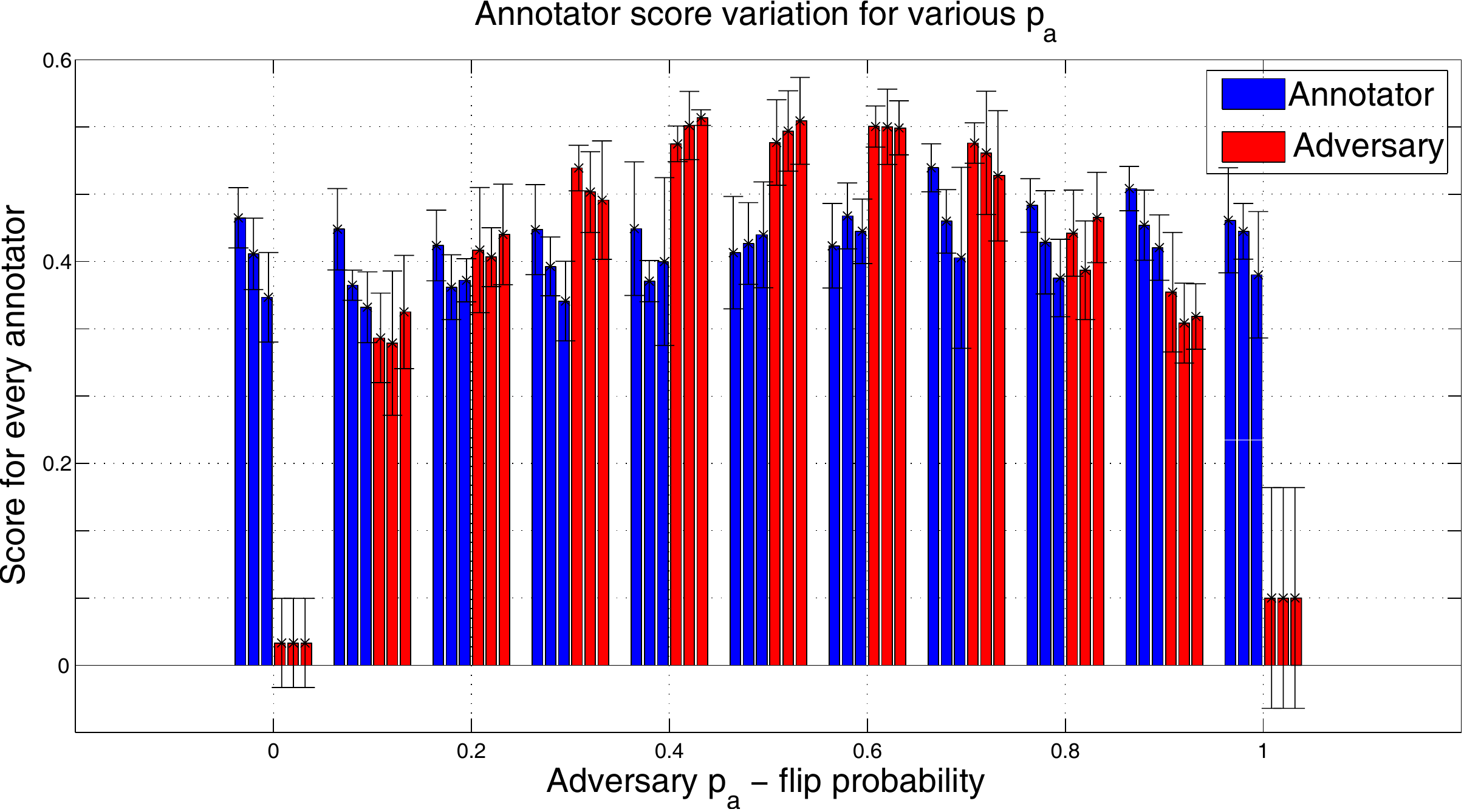}
\includegraphics[width=4.8cm]{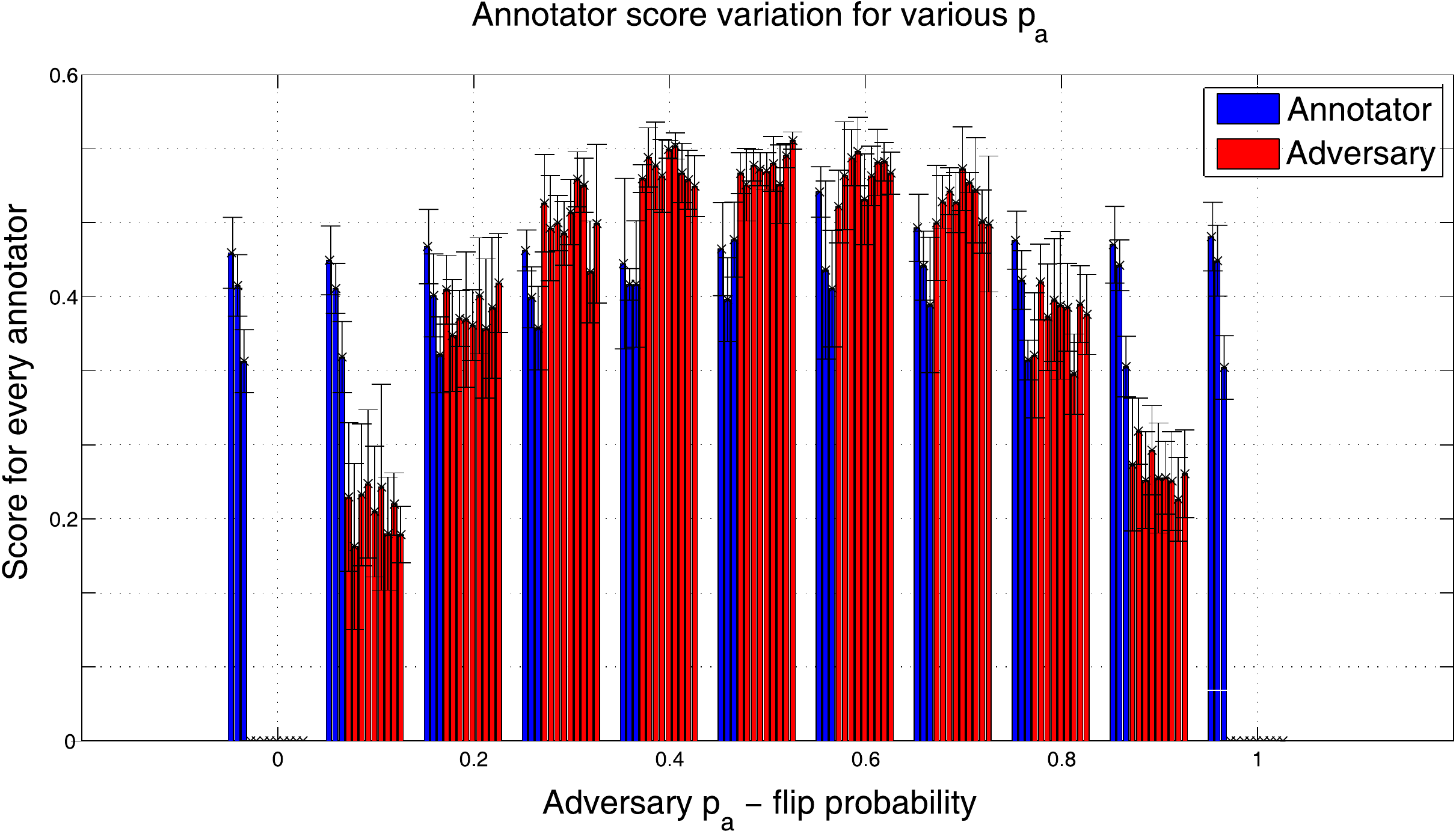}
\\
\includegraphics[width=4.8cm]{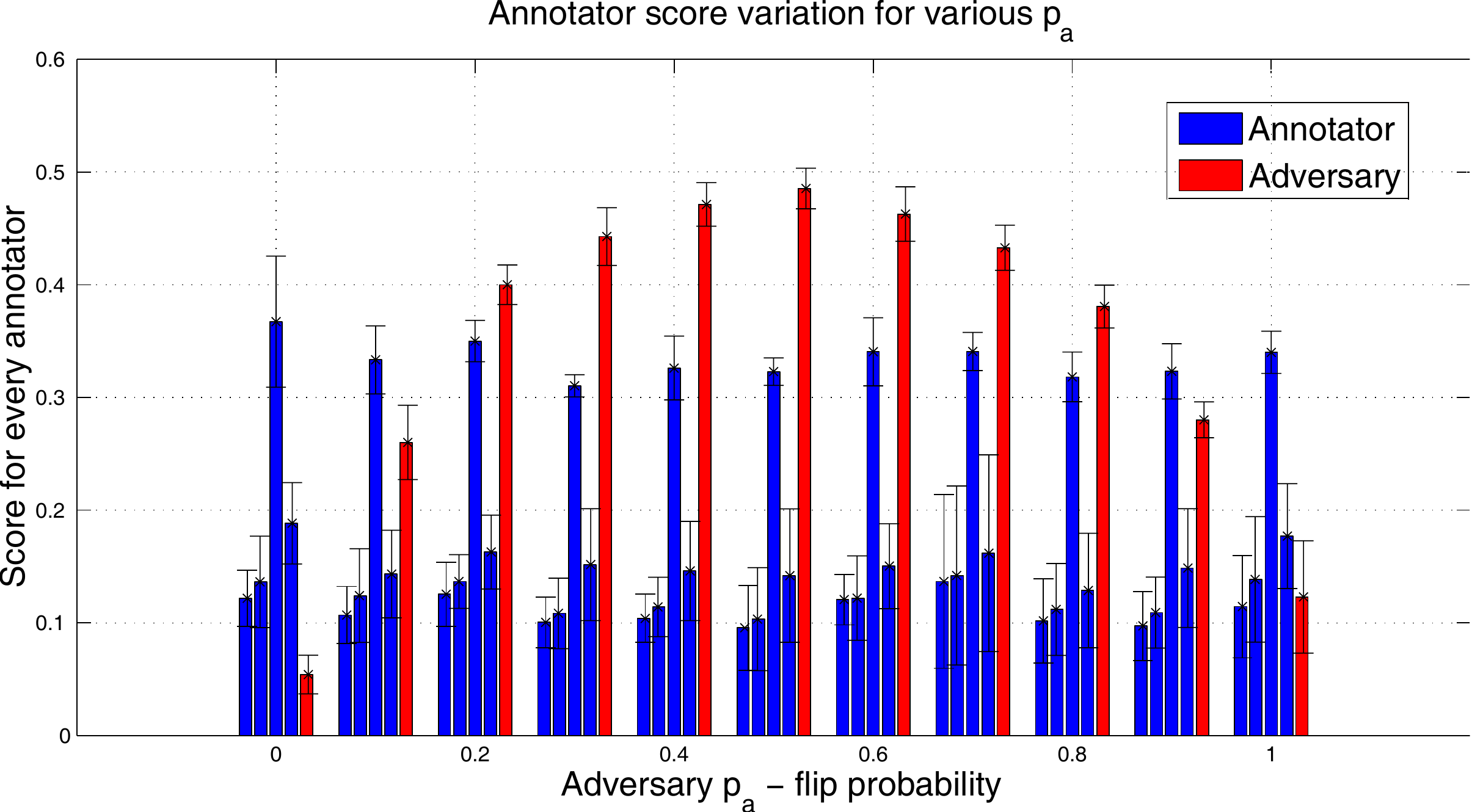}
\includegraphics[width=4.8cm]{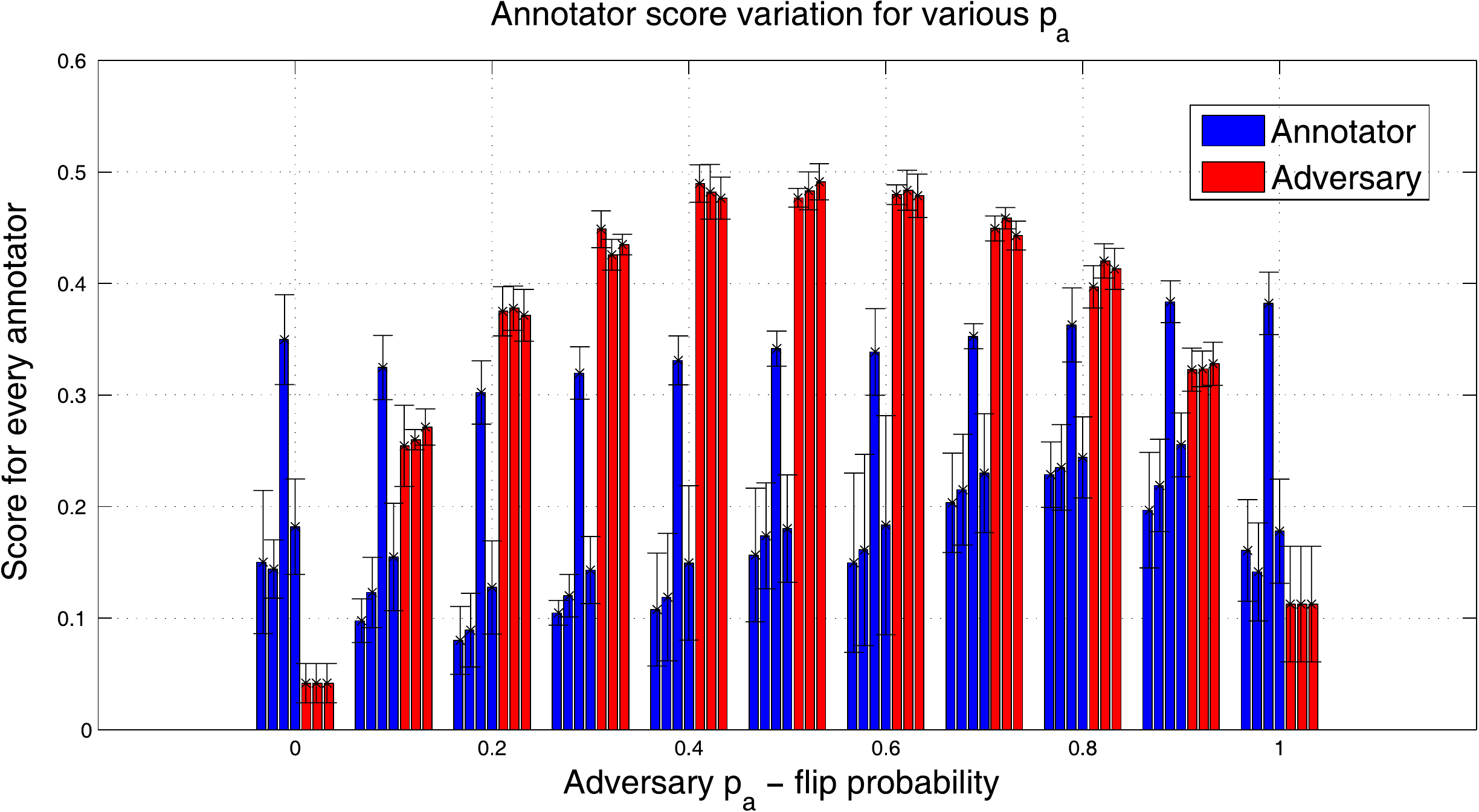}
\caption[9pt]{Scores for each annotator when 1 (left), (center), and 9 (right) adversaries are introduced for the Breast Cancer (top) and Atrial Fibrillation (bottom) datasets.}
\label{fig:adv_real}
\end{center}
\end{figure}

We start with two medical multi-annotator datasets: (1) A breast
cancer dataset where mammograms have been labeled independently
by three doctors. Ground-truth has been obtained through a biopsy, not
available to the algorithm nor the participant doctors. (2) A concept
inference dataset where non-experts read a passage of text, from
de-identified electronic medical records, to assess whether the
patient has atrial fibrillation, a type of heart
arrhythmia. Ground-truth was given by expert assessment. The datasets
contained $75$ and $1057$ datapoints respectively. For a more
controlled experiment, for both datasets we additionally generated $M$
helpful/adversarial annotators. The labels given by these were
generated as the true labels but flipped with probability $p_a$ (thus, $p_a$
effectively determines the nature of the annotator). For all
experiments we trained the multi-labeler model described in Sec.~2 to
build the distributions required by Eq.~\ref{eq:eval_annot}.

In order to illustrate the properties of the approach,
Fig.~\ref{fig:gm} shows breast cancer data points labeled by various
annotators plotted against the value of $p(y^{(k)}|\{y^{(t \backslash
  k)}\},\mathbf{x})$ for some specific annotator $k$
(Eq.~\ref{eq:eval_annot}). In Fig.~\ref{fig:gm}-center we
flipped the label of the first annotator with $p=0.4$ causing the
conditional probability of the labels given by this annotator to
decrease, as expected. In Fig.~\ref{fig:gm}-right a fourth adversarial
annotator with $p_a=0.4$ was introduced. This annotator obtains a
lower conditional probability value comparatively, indicating that it
is relatively more difficult to predict its label compared with the
original three annotators.

We additionally measured the combined impact of the number of
adversarial annotators and the strength of $p_a$. For this we defined
an {\it adversarial score} for an annotator as the sum (over data points) of the
negative log conditional probabilities defined by Eq.~1. This is
comparable with the negative log-likelihood measure frequently
utilized in learning tasks. Thus, if an annotator's labels are
difficult to infer based on the rest of the annotator labels, then the
conditional probabilities above will be generally low. The lower these
probabilities the larger the score; thus, a large score is indicative
of an annotator that is less trustworthy.

Fig.~\ref{fig:adv_real} shows this score while varying $p_a$ and the
number of helpful/adversarial annotators. We noted various properties. The
scores of the non-adversarial annotators remained fairly
constant. This indicates that the scoring scheme is robust to the
number of adversaries and the strength of $p_a$ and could be used for
detecting adversaries. The score curves for the adversaries are
symmetric about $p_a=0.5$, which is consistent with the notion that a
completely random annotator corresponds to the worst case
(independent) adversary since no information can be gained from its
labels.

\begin{figure}[t]
\begin{center}
\includegraphics[width=4.8cm]{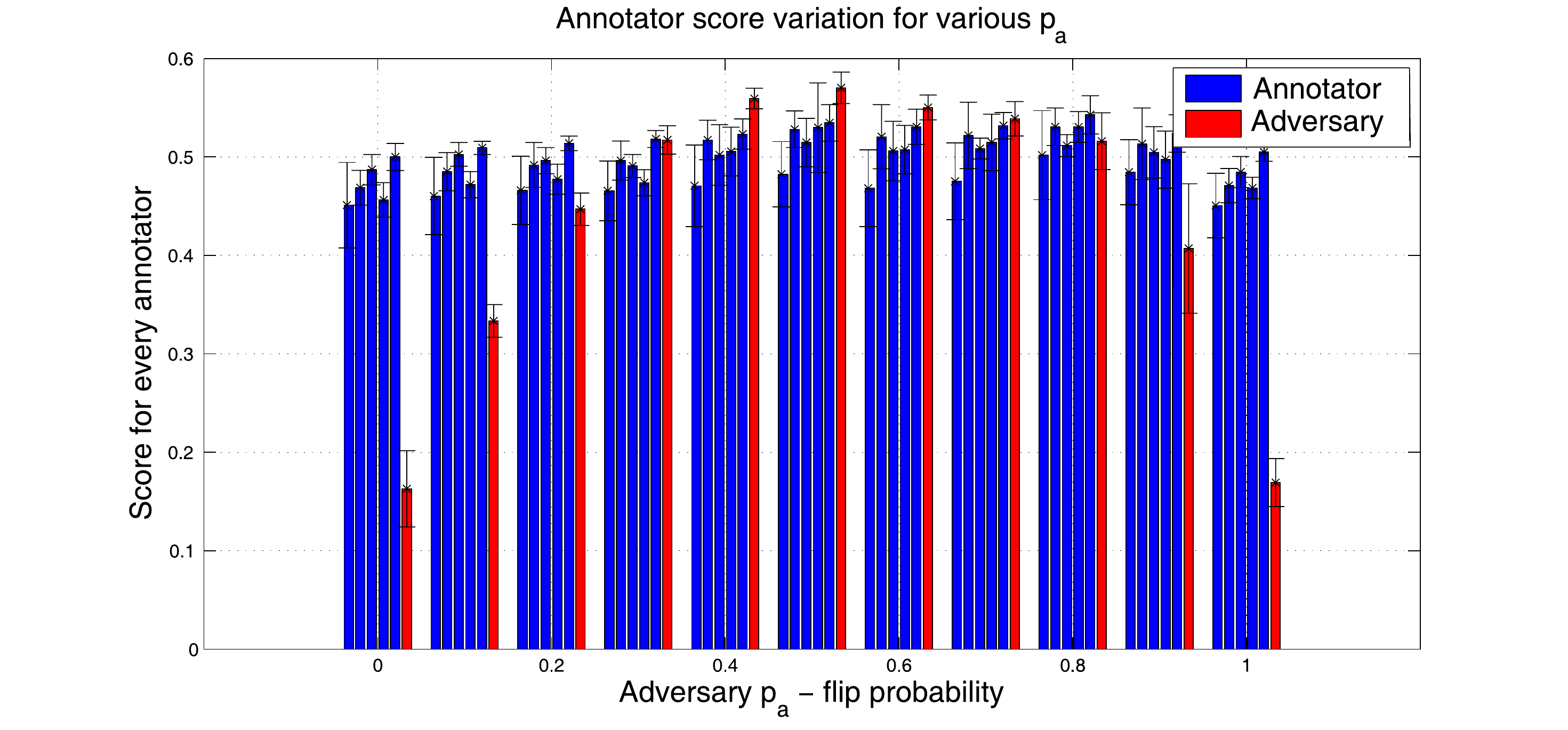}
\includegraphics[width=4.8cm]{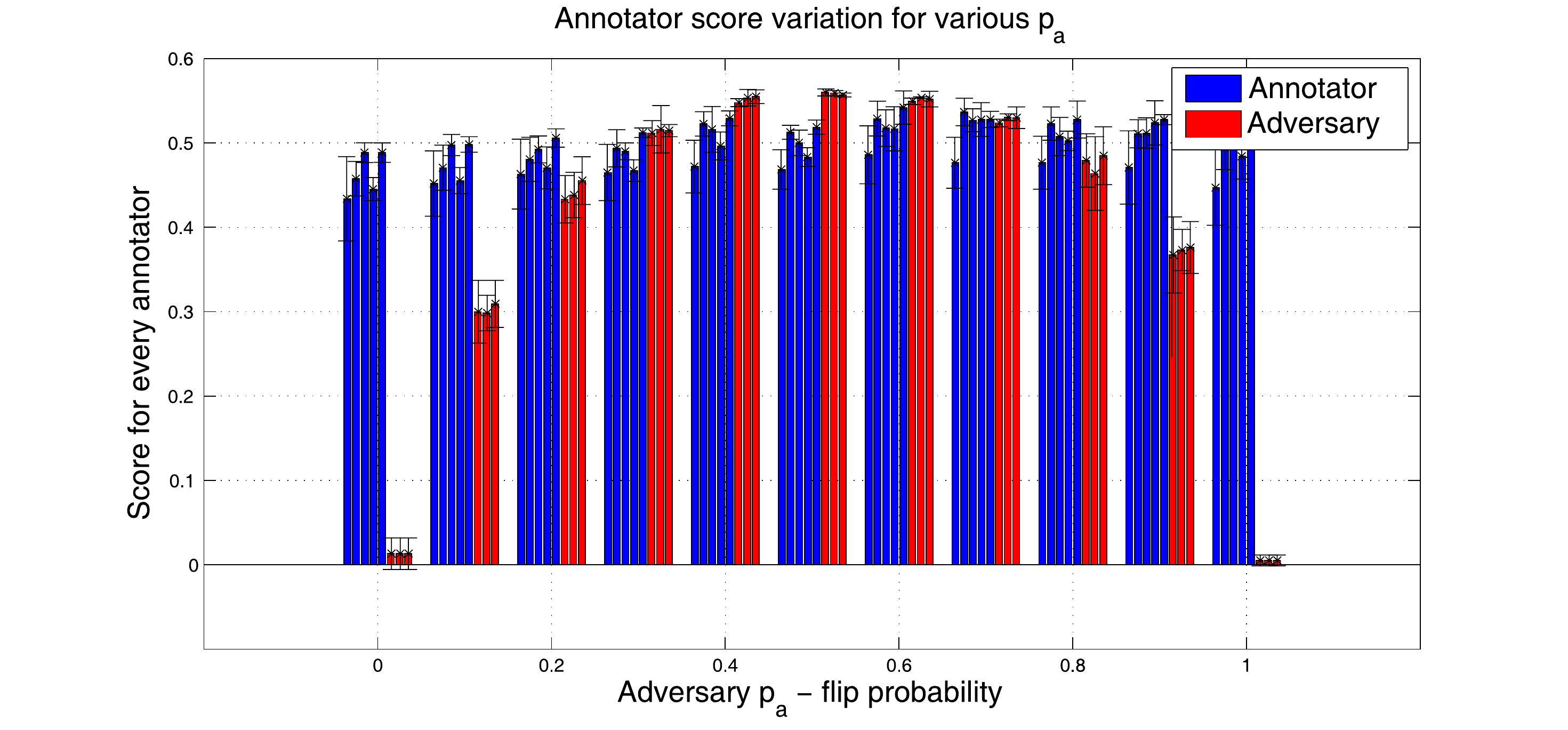}
\includegraphics[width=4.8cm]{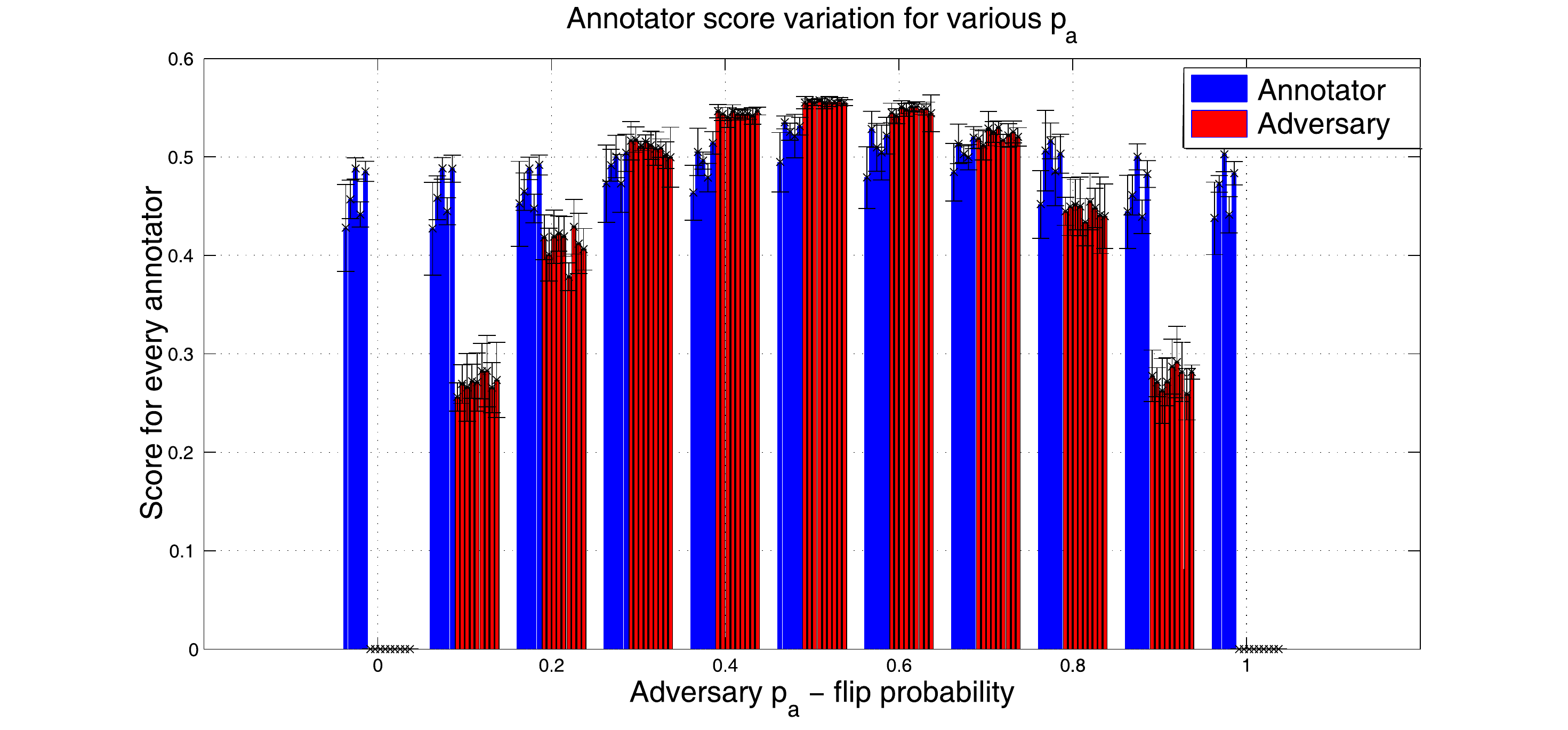}
\\
\includegraphics[width=4.8cm]{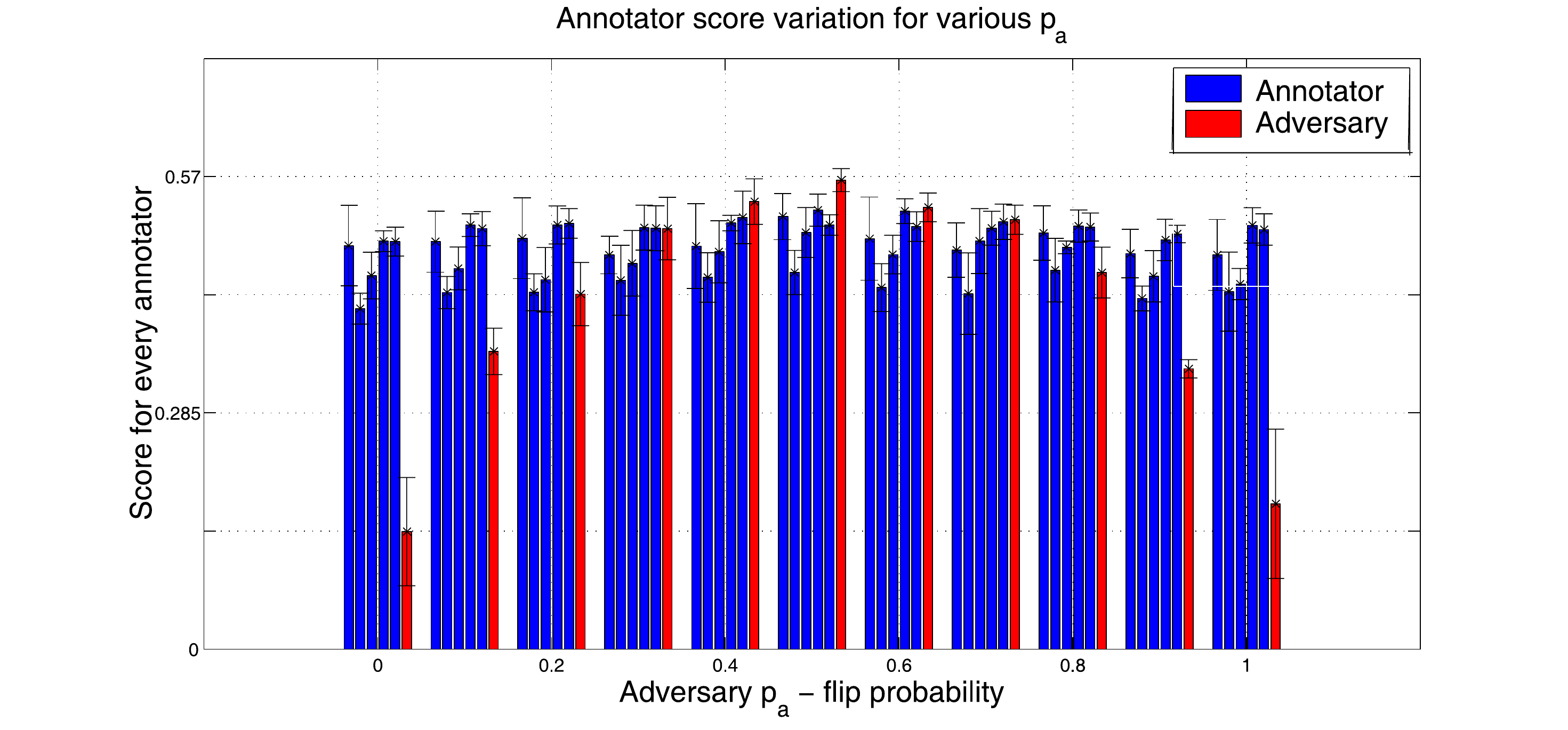}
\includegraphics[width=4.8cm]{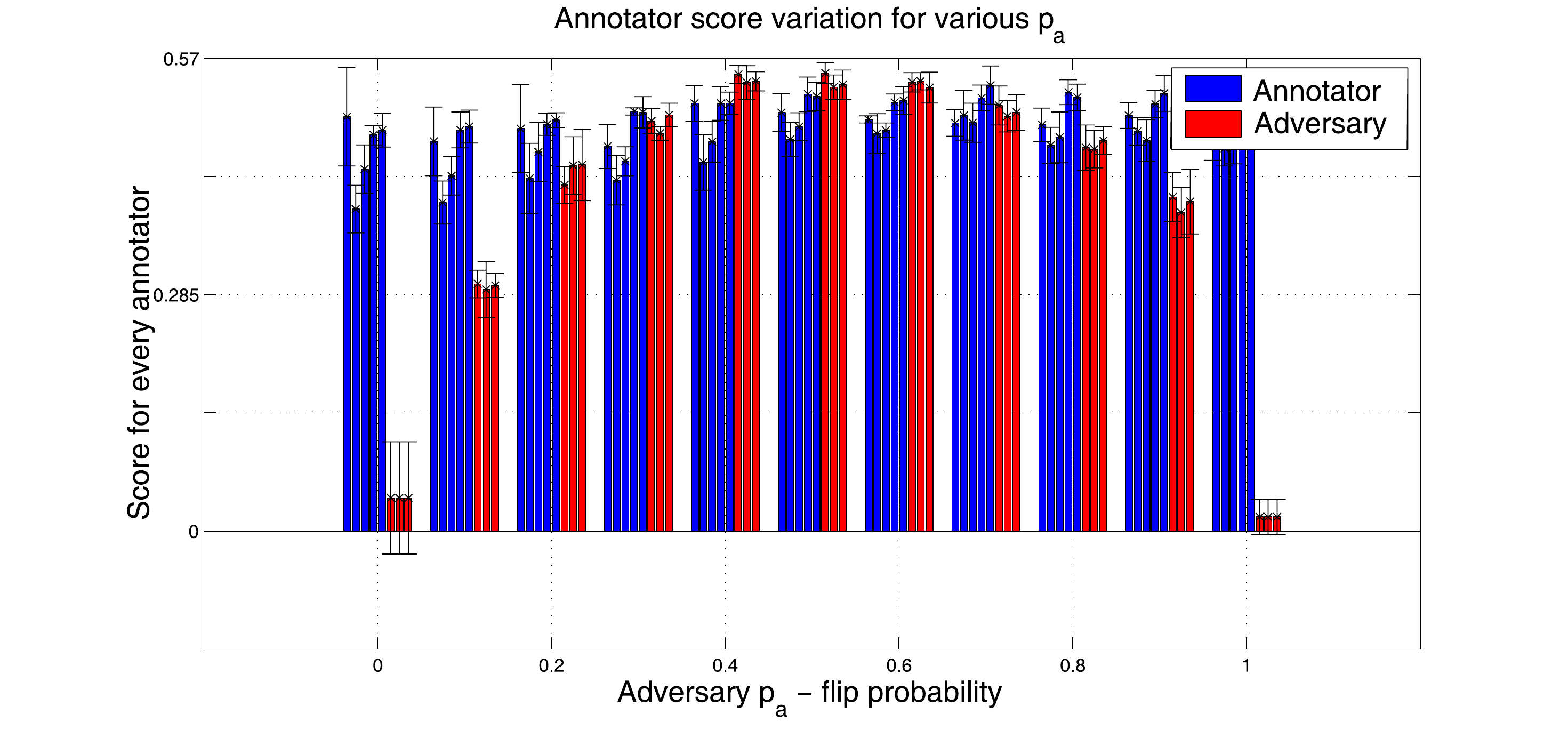}
\includegraphics[width=4.8cm]{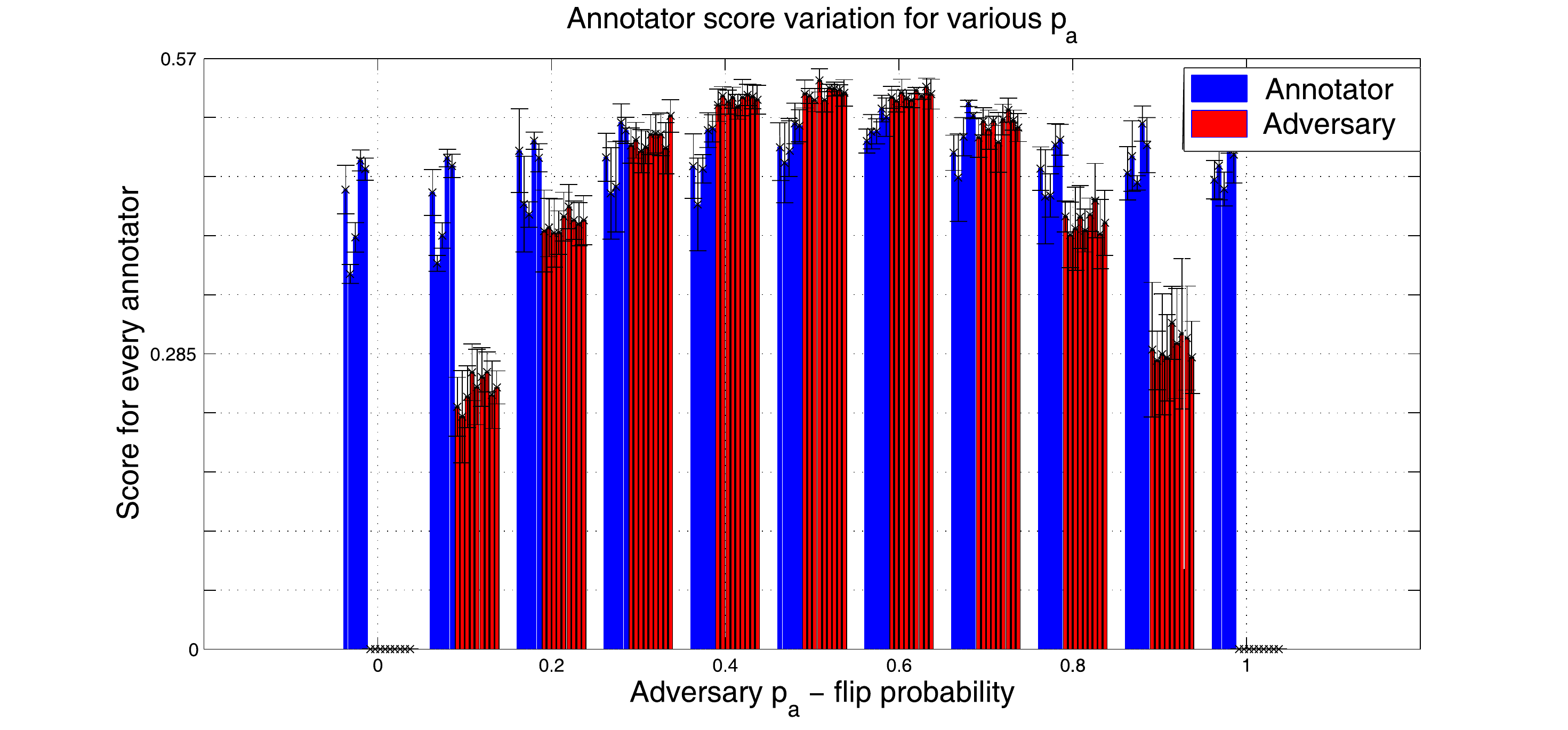}
\\
\includegraphics[width=4.8cm]{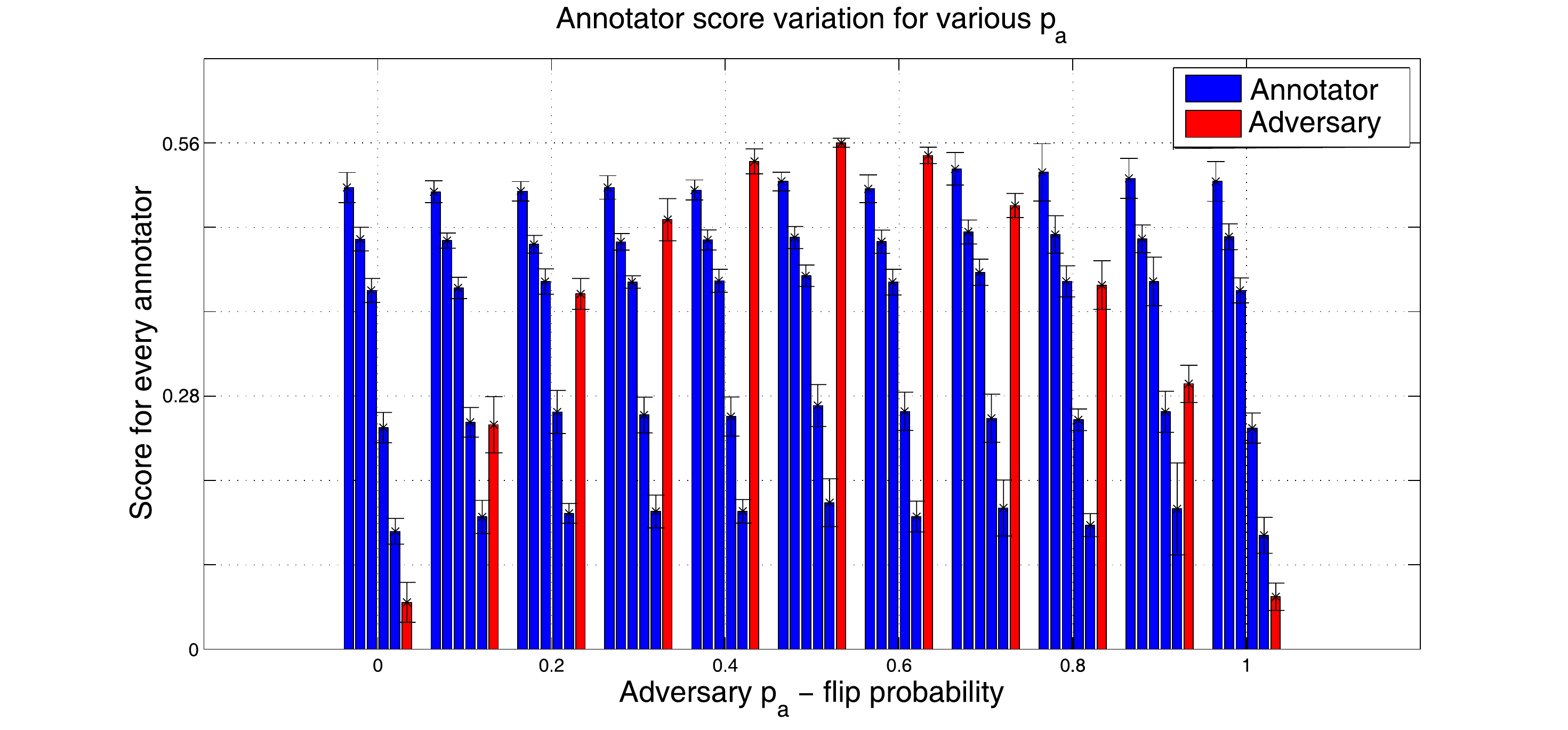}
\includegraphics[width=4.8cm]{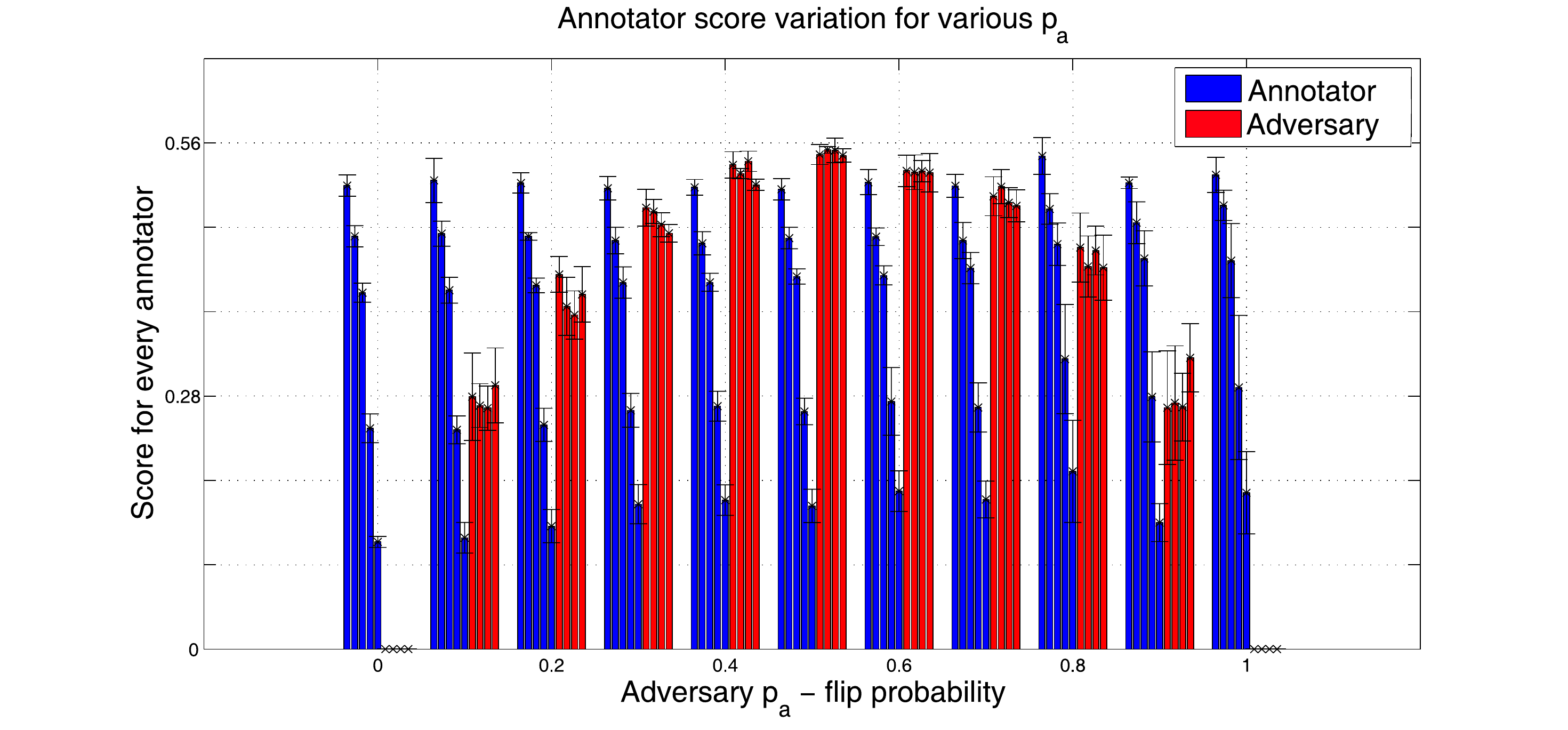}
\includegraphics[width=4.8cm]{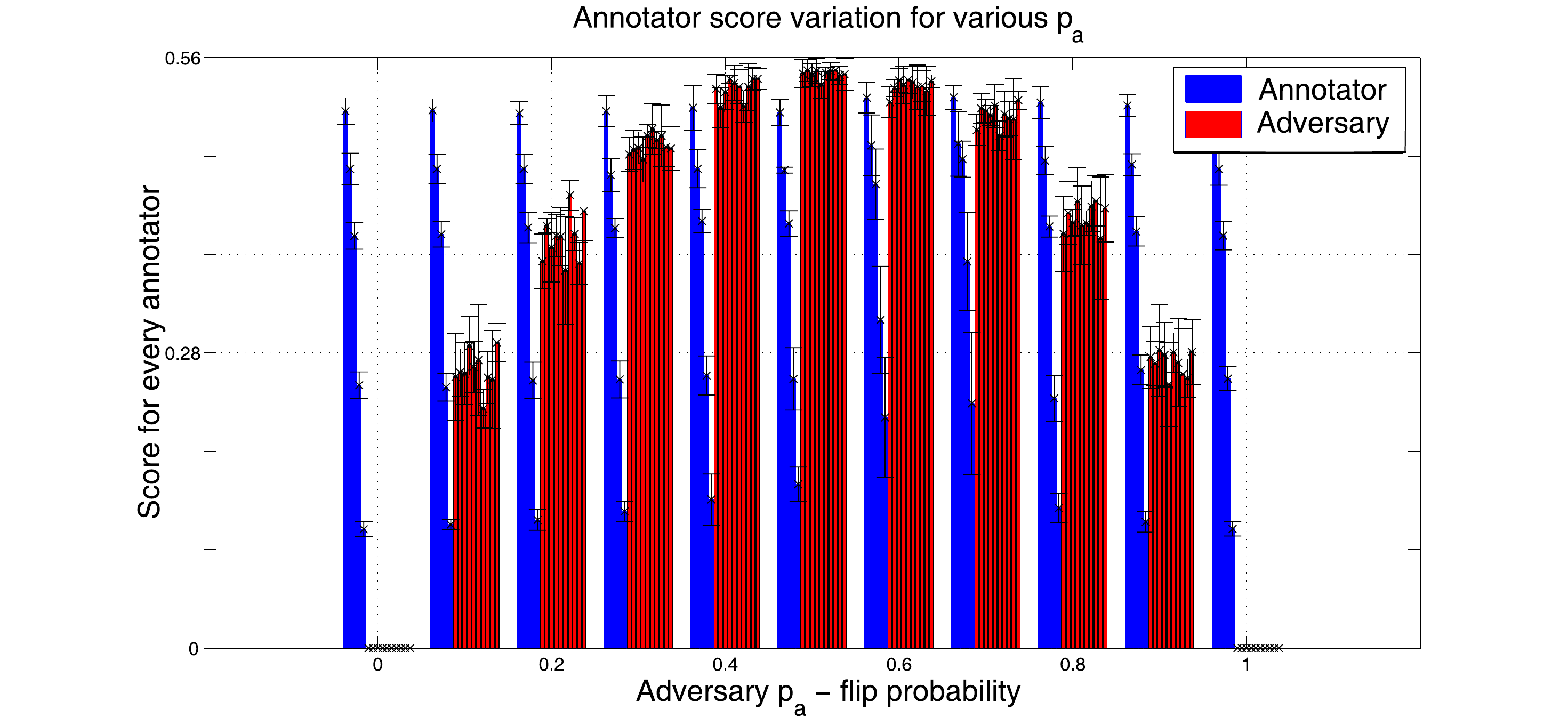}
\caption[9pt]{Scores for each annotator when 1 (left), 3 (center), and 9 (right) adversaries are introduced for the Housing (top), Ionosphere (middle), and Glass (bottom) datasets.}
\label{fig:adv_uci}
\end{center}
\end{figure}

We additionally tested this approach with various UCI datasets. Since
these datasets do not employ multiple annotators, all the
non-adversarial annotators (blue) were generated at random based on
the true label, while the adversarial annotators (red) were generated
as before. Results, shown in Fig.~\ref{fig:adv_uci} are consistent
with the previous experiments and further confirm the benefits of the approach in a more controlled setting.

\begin{figure}
\begin{center}
\includegraphics[width=4.8cm]{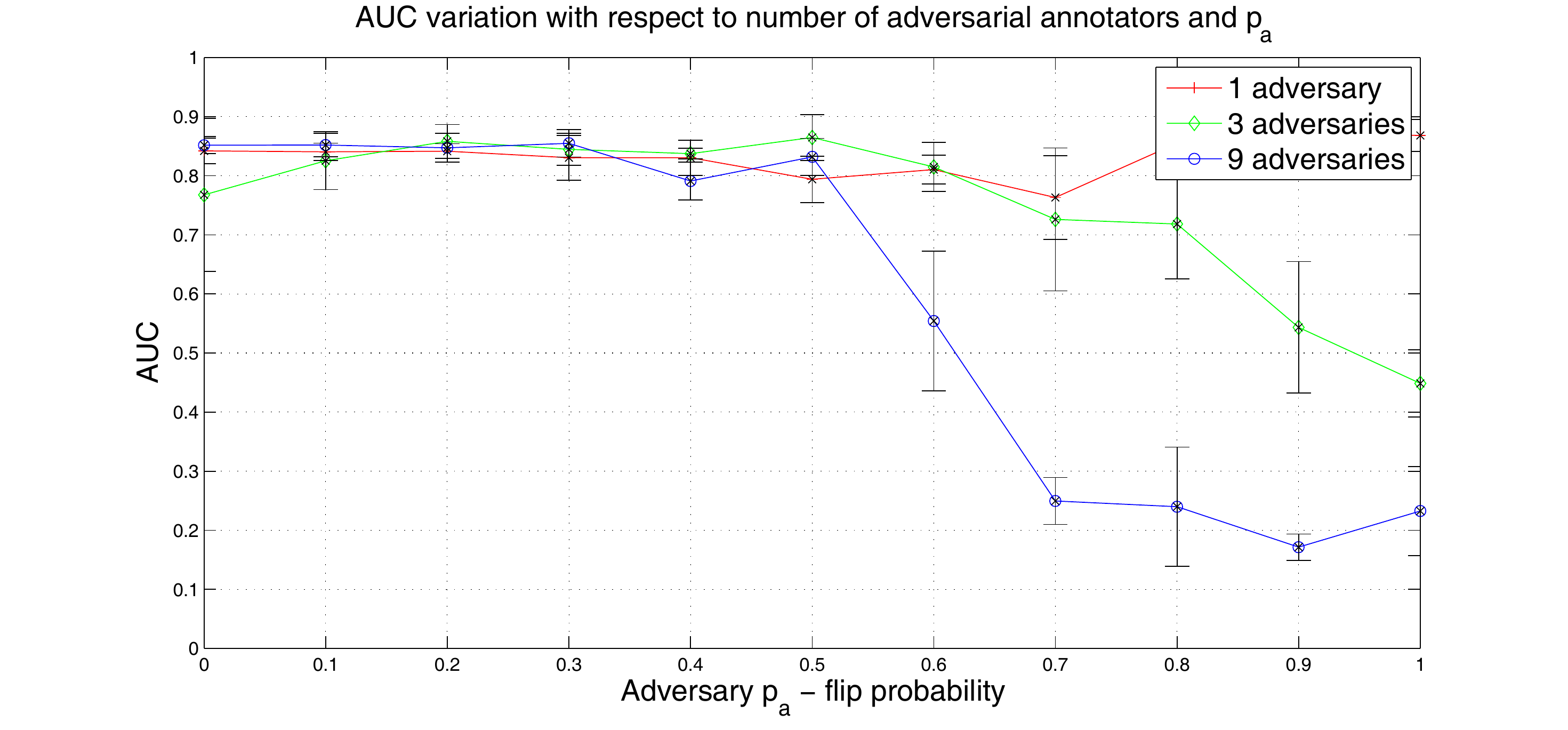}
\includegraphics[width=4.8cm]{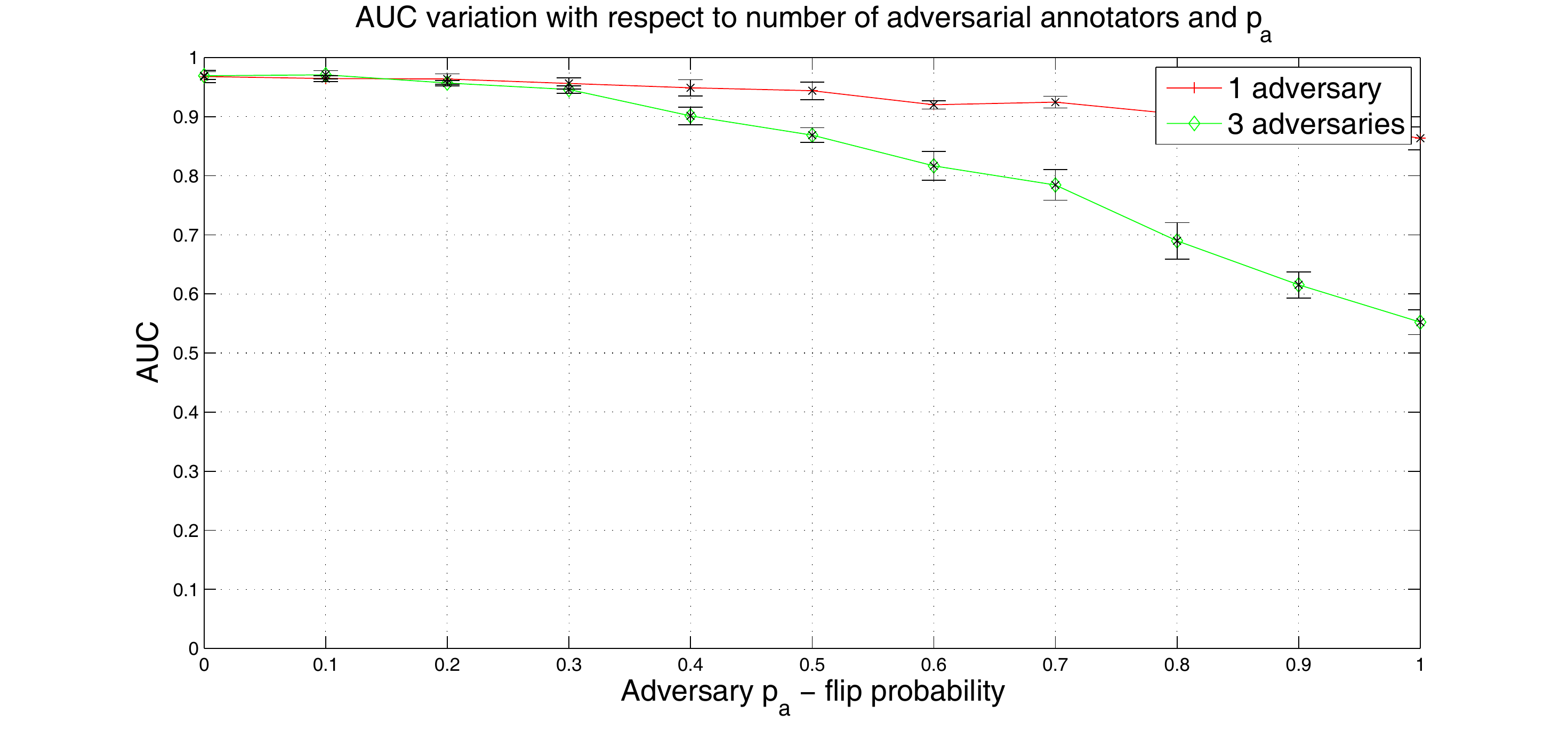}
\caption[9pt]{AUC for various $p_a$ values and varying number of helpful/adversarial annotators for the Breast Cancer and Atrial Fibrillation datasets.}
\label{fig:auc_real}
\end{center}
\end{figure}

\begin{figure}
\begin{center}
\includegraphics[width=4.8cm]{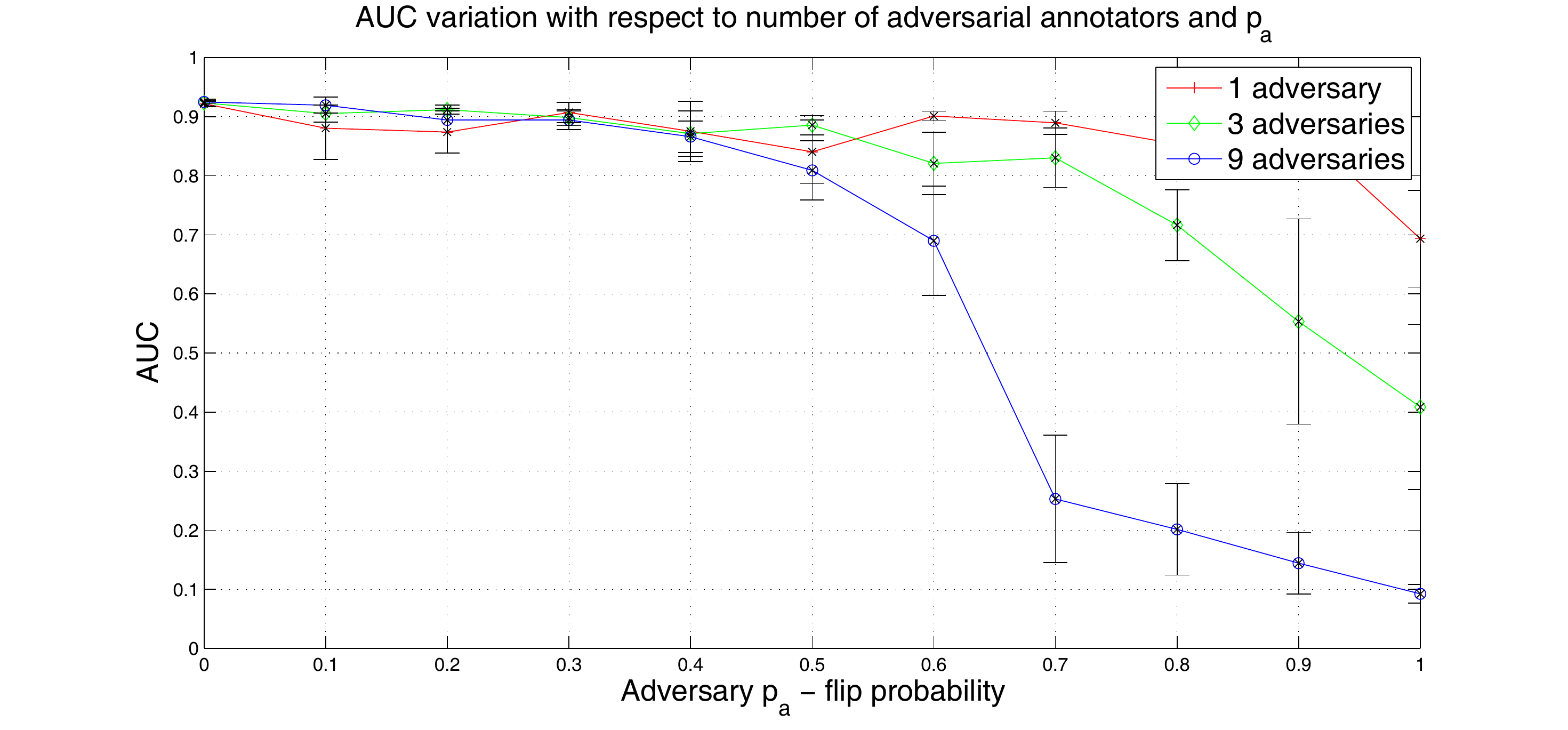}
\includegraphics[width=4.8cm]{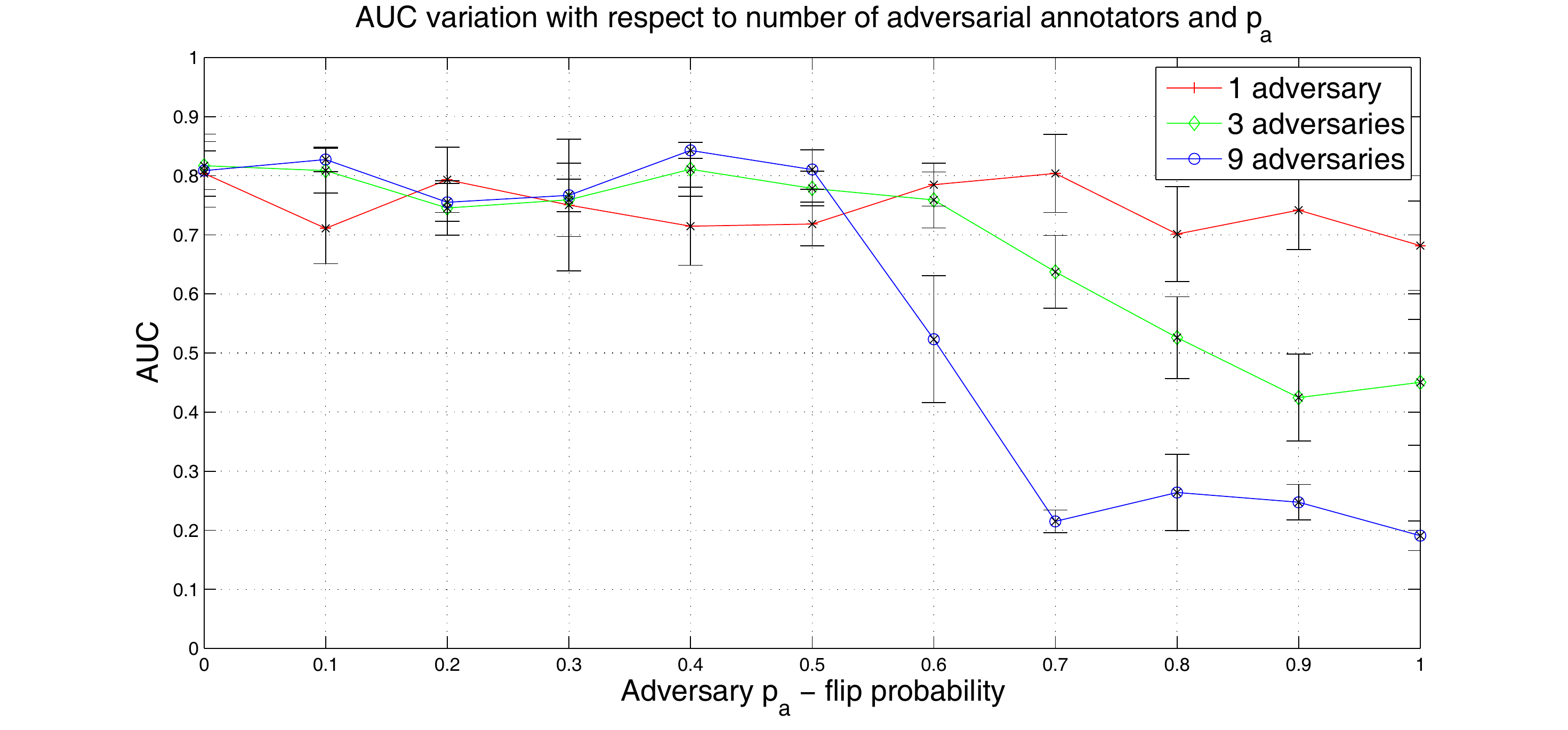}
\includegraphics[width=4.8cm]{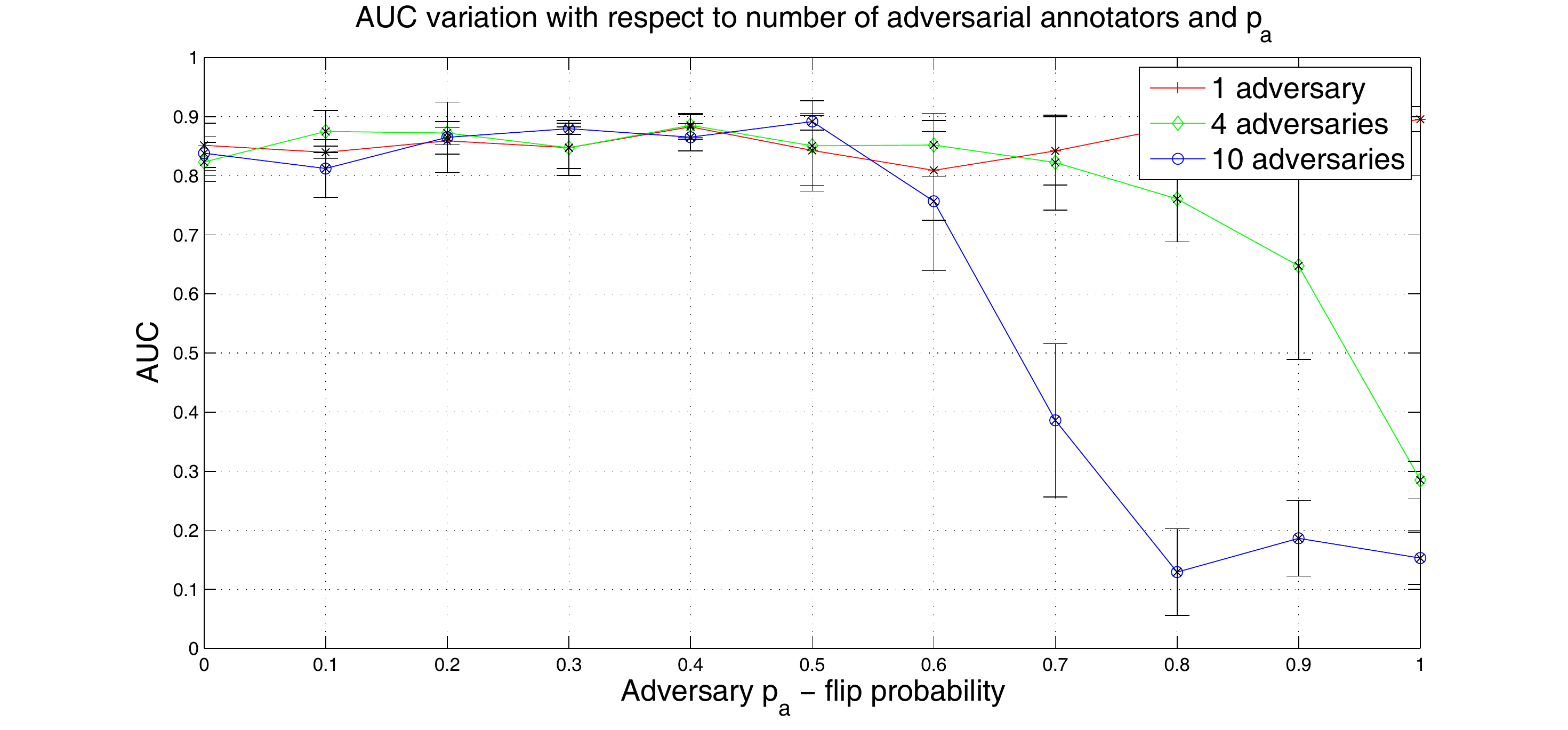}
\caption[9pt]{AUC for various  $p_a$ values and varying number of helpful/adversarial annotators for the Housing, Ionosphere, and Glass datasets.}
\label{fig:auc_uci}
\end{center}
\end{figure}

Finally, Figs.~\ref{fig:auc_real}-\ref{fig:auc_uci} show how the
effect of adversarial annotators and the value of $p_a$ can impact the
performance (AUC) of the model. We can observe that the AUC decreases
considerably only when a large proportion of the adversarial annotators
consistently flip the correct label. This indicates that the model can
remain accurate under a wide range of adversarial influence.

